# Sub-monolayer Biolasers: Lower Gain, Higher Sensitivity

Chaoyang Gong[1,3,7], Xi Yang[1,2,7], Shui-Jing Tang[2,7], Qian-Qian Zhang[1,7], Yanqiong Wang[1], Yi-Ling Liu[1], Yu-Cheng Chen[3], Gang-Ding Peng[5], Xudong Fan[6], Yun-Feng Xiao[2,*], Yun-Jiang Rao[1,4,*], and Yuan Gong[1,*]

[1] Key Laboratory of Optical Fiber Sensing and Communications (Ministry of Education of China), School of Information and Communication Engineering, University of Electronic Science and Technology of China, Chengdu, Sichuan 611731, People's Republic of China.

[2] State Key Laboratory for Mesoscopic Physics and Frontiers Science Center for Nano-optoelectronics, School of Physics, Peking University, Beijing 100871, People's Republic of China.

[3] School of Electrical and Electronic Engineering, Nanyang Technological University, Singapore 639798, Singapore.

[4] Research Center for Optical Fiber Sensing, Zhejiang Laboratory, Hangzhou, Zhejiang 310000, People's Republic of China.

[5] School of Electrical Engineering and Telecommunications, University of New South Wales, Sydney, NSW 2052, Australia.

[6] Department of Biomedical Engineering, University of Michigan, Ann Arbor, Michigan 48109, USA.

[7] These authors contributed equally: Chaoyang Gong, Xi Yang, Shui-Jing Tang, and Qian-Qian Zhang.

*E-mails: yfxiao@pku.edu.cn; yjrao@uestc.edu.cn; ygong@uestc.edu.cn

**Abstract**

Biomarker detection is the key to identifying health risks. However, designing sensitive biosensors in a single-use mode for disease diagnosis remains a major challenge. Here, we report sub-monolayer biolasers with remarkable repeatability for ultrasensitive and disposable biomarker detection. The biolaser sensors are designed by employing the telecom optical fibers as distributed optical microcavities and pushing the gain molecules down to the sub-monolayer level. We observe a status transition from the monolayer biolaser to the sub-monolayer biolaser by tuning the specific conjugation. By reducing the fluorophores down to the threshold density ($\sim 3.2 \times 10^{-13}$ mol/cm$^2$), we demonstrate an ultimate sensitivity of sub-monolayer biolaser with six orders of magnitude enhancement compared with the monolayer biolasers. We further achieved ultrasensitive immunoassay for Parkinson's disease biomarker, alpha-synuclein, with a lower limit of detection of 0.32 pM in serum. This biosensor with massive fabrication capability at ultralow cost provides a general method for the ultrasensitive disposable biodetection of disease biomarkers.

# 1. Introduction

Early detection of diseases like cancer and dementia before they manifest serious, irreversible symptoms is of considerable public health importance, which can help to reduce morbidity and mortality[1-4]. At the early stage of a disease, it is extremely difficult to precisely estimate the extremely low concentrations of the biomarkers[5-9]. Optical biosensors, which amplify the weak biological signal by enhancing the light-matter interaction, are one mainstream technology for sensitive biomarker detection[10-13]. To date, various types of optical methods based on interferometers[14,15], surface plasmon resonance (SPR)[16-22], surface-enhanced Raman scattering (SERS)[23-27], and optical microcavities[28-35] have been developed to break the record of the lower limit of detection (LOD). Their performance, on the other hand, relies highly on meticulous design and precise fabrication, making high-throughput production of disposable diagnostic devices difficult[10,36,37]. Due to the amplification effect, even a minor fabrication error can cause considerable deviations in test results and deteriorates the disposability[38-41]. This is particularly the case with ultrasensitive biosensors. Micro- and nano-interferometers rely on precision micromachining facilities, such as femtosecond lasers[42], focused ion beams[43], or electron beams[44]. Surface plasmon resonance biosensors require thin film deposition with precise thickness at the nanometer scale[45]. The SERS signals are strongly dependent on the properties of nanoparticles and substrates[46]. Optical microcavities have evolved as a powerful platform for amplifying optical signals with strong cavity feedback over the last two decades, and it has been widely used for biomolecule[35,47,48] and cell analysis[49-51]. The strong dependence on delicate fabrication procedures and the essential coupling requirement, however, are highly undesired for single-use biosensors.

Here, we propose the concept of sub-monolayer biolasers to bridge the gap between sensitivity and disposability of microcavities. The sub-monolayer biolasers were mass-produced at negligible cost using optical fiber microcavities that were distributed across an extraordinary length of 10 km and had ultrahigh Q-factors of $10^6$ (Fig. 1a). In striking contrast to the passive microcavity, the pump and detection of sub-monolayer biolasers can be conveniently performed by free-space optics, which release the dependence on the critical waveguide coupling and more importantly, make available the disposable biosensors with ultrahigh sensitivity. As illustrated in Figs. 1b and 1c, by pushing the gain molecules down to the threshold density (~$3.2 \times 10^{-13}$ mol/cm$^2$), the sub-monolayer biolaser demonstrates a considerable enhancement in sensitivity over six orders of magnitude compared to the monolayer biolasers. The sub-monolayer biolaser combines the reproducible fiber microcavity with the stimulated amplification effect of a sub-molecular layer gain, resulting in ultrahigh sensitivity and disposability. The sub-monolayer biolaser was further employed to detect a PD biomarker in serum with a lower LOD of 0.32 pM. We envision that the ultrahigh sensitivity and massive disposability of microlaser biosensors could enable the cost-effective and early diagnosis of major diseases.

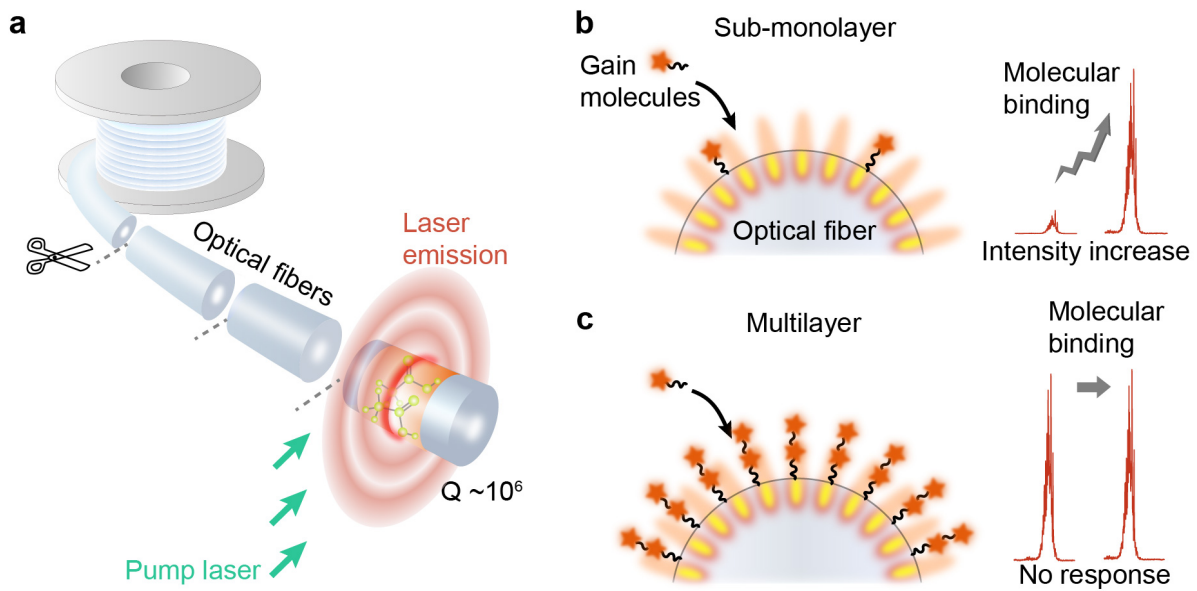

**Figure 1** Schematic illustration of the sub-monolayer biolaser. (**a**) Massive production of the sub-monolayer biolaser using optical fiber. Conceptual illustration of the sensitivity enhancement in (**b**) the sub-monolayer biolaser compared to (**c**) the monolayer or multilayer biolasers.

## 2. Results

### 2.1 Conceptual demonstration of sub-monolayer biolasers

We demonstrated a novel microcavity biolaser based on the sparse and specific conjugation of molecules on the optical fiber microcavity and observed for the first time, the transit from a monolayer biolaser into the sub-monolayer biolaser (Fig. 2). The comprehensive optical fiber (SMF-28e, Corning) was employed for the conceptual demonstration of the sub-monolayer biolasers. The optical fibers were biotinylated and then conjugated with streptavidin-Cy3 (Sav-Cy3) molecules (Supplementary Fig. 1) that served as laser gain. The bright fluorescence image confirms the successful conjugation of the Cy3 molecules (Inset of Fig. 2a). The optical fiber microcavities show a high Q-factor exceeding $10^6$ in the aqueous environment (Supplementary Fig. 2), which enables the strong light-matter interaction on the surface. The sharp peaks in the emission spectrum (Fig. 2a) and the threshold behavior (Supplementary Fig. 3) confirm the successful generation of laser emission from a sub-monolayer of gain molecules. Compared with the fiber taper coupling in the passive microcavity sensors, the lateral laser pump and emission collection in free space greatly improve the reproducibility for disposable use. The penetration depth of the evanescent wave is over 10 times the thickness of the molecular layer (Supplementary Fig. 4). Owing to the high localization of molecules on the fiber surface, all the gain molecules can participate in lasing, which is different from the homogeneous distribution of molecules in optofluidic lasers with a liquid core optical ring resonator (LCORR)[52-54] so that an ultrahigh sensitivity and an ultralow fluorescence background can be achieved. The laser emission is strongly dependent on the surface density of the Cy3 molecules, which can be specifically controlled by the biotin (Supplementary Fig. 1). The laser threshold increases from 0.09 mJ/mm$^2$ to 0.22 mJ/mm$^2$ (Supplementary Fig. 3) when the surface density decreases from $1.8 \times 10^{-12}$ mol/cm$^2$ to $8.2 \times 10^{-13}$ mol/cm$^2$ (Point A to Point B in Supplementary Fig. 5). Note that we tried to minimize the number of gain molecules to maximize the sensitivity of biodetection, rather than to optimize the laser threshold.

We explored the ultimate limit of the ultrathin layer to lase by reducing the surface density of gain molecules (Fig. 2b), according to the specific binding mechanism (Supplementary Fig. 1). Firstly, monolayer biolasers were demonstrated and the laser emission remained stable when gradually decreasing the biotin down to 0.4 mM (the saturated zone in Fig. 2b). Note that the residual molecules were washed away to enable specific binding of biomolecules, therefore, the upper limit is a saturated monolayer. Then, a turning point that indicates a status transition from monolayer to sub-monolayer biolaser was observed with a biotin concentration of around 0.4 mM. According to the analysis in Supplementary Fig. 5, the surface density at this turning point is about $1.8 \times 10^{-12}$ mol/cm$^2$.

Subsequently, the laser intensity decreases linearly with biotin in the log-log scale (the gradient zone in Fig. 2b), which defines the concept of sub-monolayer biolaser.

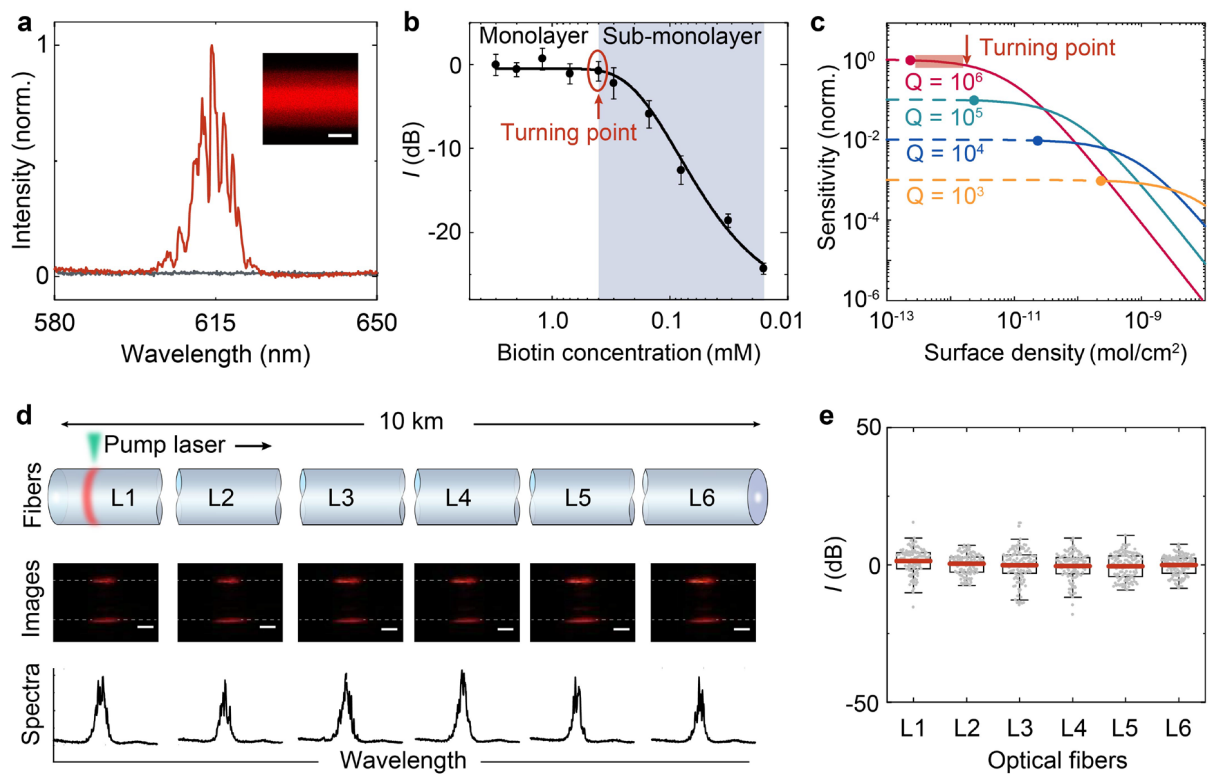

**Figure 2 Characterization of the sub-monolayer biolasers.** (**a**) The emission spectra. Red curve, pump above the threshold at 0.46 mJ/mm$^2$; gray curve, pump below the threshold at 0.06 mJ/mm$^2$. Inset, fluorescence image of the sub-monolayer biolaser. Scale bar, 50 μm. (**b**) The transition from the monolayer biolaser to the sub-monolayer biolaser by reducing the conjugation of biotin molecules. Each error bar is calculated from more than 60 lasers and denotes the 95% confidence interval. (**c**) Theoretical sensitivity as a function of Cy3 surface density. The dots denote the threshold density for lasing, while the dashed curves correspond to the cases below the lasing threshold. (**d**) Laser images and spectra of disposable sub-monolayer biolasers, with six segments of optical fibers (L1 to L6) randomly picked from a 10 km spool. Scale bar, 50 μm (**e**) Statistics of the laser intensity. The data was collected from 736 lasers on six segments of fibers. Blue bars, the average intensity of each fiber.

## 2.2 Sensitivity analysis of the sub-monolayer biolasers

In order to illustrate the sensing mechanism of sub-monolayer biolasers, we developed a theoretical model (Supplementary Eqs. 1 to 11) and calculated the sensitivity (Fig. 2c). The numerical results indicate that the sensitivity continues to increase with a lower surface density and a higher Q-factor. Thanks to the high Q-factor exceeding $10^6$, an ultrahigh sensitivity can be demonstrated when the surface density was pushed down to lower than the turning point ($1.8 \times 10^{-12}$ mol/cm$^2$, Fig. 2b). In contrast, the linear fluorescence emission shows no improvement in sensitivity with less fluorescent molecules (Supplementary Eqs. 11 to 12). This phenomenon enables us to achieve an ultrahigh sensitivity by decreasing the surface density of biomarker-related gain molecules.

**2.3 Disposable test of sub-monolayer biolasers**

Telecom optical fibers were proposed by Charles Kao half a century ago and have been revolutionized in design, material, and drawing process. Currently, the mature fiber draw tower technology enables the highly precise control of the fiber geometry, including the cladding diameter and non-circularity, and realizes the cost-effective fabrication of > 50,000 meters of optical fibers with one preform. Silica optical fibers can be considered as a series of distributed cylindrical optical microcavities that supports WGMs. Their high reproducibility and ease of fabrication with ultralow cost make optical fiber an ideal platform for disposable use.

We investigated the disposability of the sub-monolayer biolasers that were fabricated with different segments of optical fibers randomly selected from a 10 km spool. The Q-factors of the distributed microcavities at different locations were tested in the buffer to be $1.2 \times 10^6$ (Supplementary Fig. 2c). Thanks to the uniform optical properties along the fiber, the lasing threshold of the sub-monolayer biolasers is below 1 mJ/mm$^2$, which is within the acceptable range for biomedical samples (Supplementary Fig. 6). We further tested 736 biolasers that were distributed on six segments of optical fibers (L1 to L6) by linearly scanning the pump laser and simultaneously recording the laser pattern and spectra (Fig. 2d). The emission properties of these biolasers are similar, owing to the uniform geometry and surface properties of the optical fibers. In order to quantify the laser emission of the sub-monolayer biolasers, we calculated the relative laser intensity in decibels (dB) (see Methods for details), which agrees well with a normal distribution (Supplementary Fig. 7). Since the pump laser scans along with the optical fiber with a step of 250 μm, each sub-monolayer biolaser on the pump location was regarded as a sensing element for disposable uses. A considerable number of sub-monolayer biolasers can be measured in a short time, which enables statistical analysis of sub-monolayer biolasers. To reduce the random variations in sensing, we employed the average laser intensity as the sensing indicator throughout the experiment, showing a small variation of 2% (Fig. 2e). This result confirms the capability of the sub-monolayer biolasers for disposable use. A total of 5163 biolasers were employed as the disposable sensing elements to demonstrate their massive disposability, including those for threshold tests, conceptual demonstration, and biomarker detection. The disposable sub-monolayer biolasers are inherently safe and free of recalibration. They also enable parallel and high throughput tests.

## 2.3 The ultimate sensitivity of the sub-monolayer laser-based biosensors

We experimentally demonstrated an ultrasensitive laser-based biosensor by reducing the number of gain molecules to the sub-monolayer level. The biotin-avidin interaction was employed as a demonstrative model due to its high affinity and specificity. This specific molecular conjugation has been widely exploited in the mainstream clinical methods including western blotting (WB) and enzyme-linked immunosorbent assay (ELISA). We fabricated three types of biolasers on optical fiber, i.e., sub-monolayer biolaser (Fig. 3a), monolayer biolaser (Fig. 3d), and multilayer biolasers (Fig. 3g). The sub-monolayer and monolayer biolasers were distinguished by controlling the biotin concentrations, while the multilayer biolasers were formed by adding multilayers of Cy3 molecules using streptavidin bridging molecules[55]. The details of the fabrication process are described in Methods.

We compared the sensing performance of the three types of biolasers by employing the competitive binding assay of avidin as an example. The analyte (avidin) competes with the gain (Sav-Cy3) molecules for the binding sites (biotin) on fiber. With a higher concentration of the avidin, more binding sites (biotin) were occupied by avidin so that fewer Sav-Cy3 molecules were conjugated to generate a lower laser intensity. In the sub-monolayer biolaser, an ultrahigh sensitivity was demonstrated, owing to the significant decrease of gain molecules. An intensity decrease of about 40% was observed between the blank and 100 pM (Fig. 3b). The statistical distribution started to shift towards the lower intensity when the avidin concentration was above 10 pM (Fig. 3c). In the monolayer biolasers, the intensity decreases slightly (~ 7%) at 100 pM of avidin (Fig. 3e), and a horizontal shift in statistical distribution can be distinguished at 1000 pM (Fig. 3f). The multilayer biolasers have a relatively large number of gain molecules on fiber, therefore, no significant change in statistical distribution can be observed and the slight change of gain molecules caused by avidin binding is ignorable (Fig. 3h). Even when the avidin concentration reaches 1000 pM, the avidin binding effect is still indistinguishable in the laser emission (Fig. 3i). These results indicate a significant enhancement of sensitivity by the sub-monolayer biolasers.

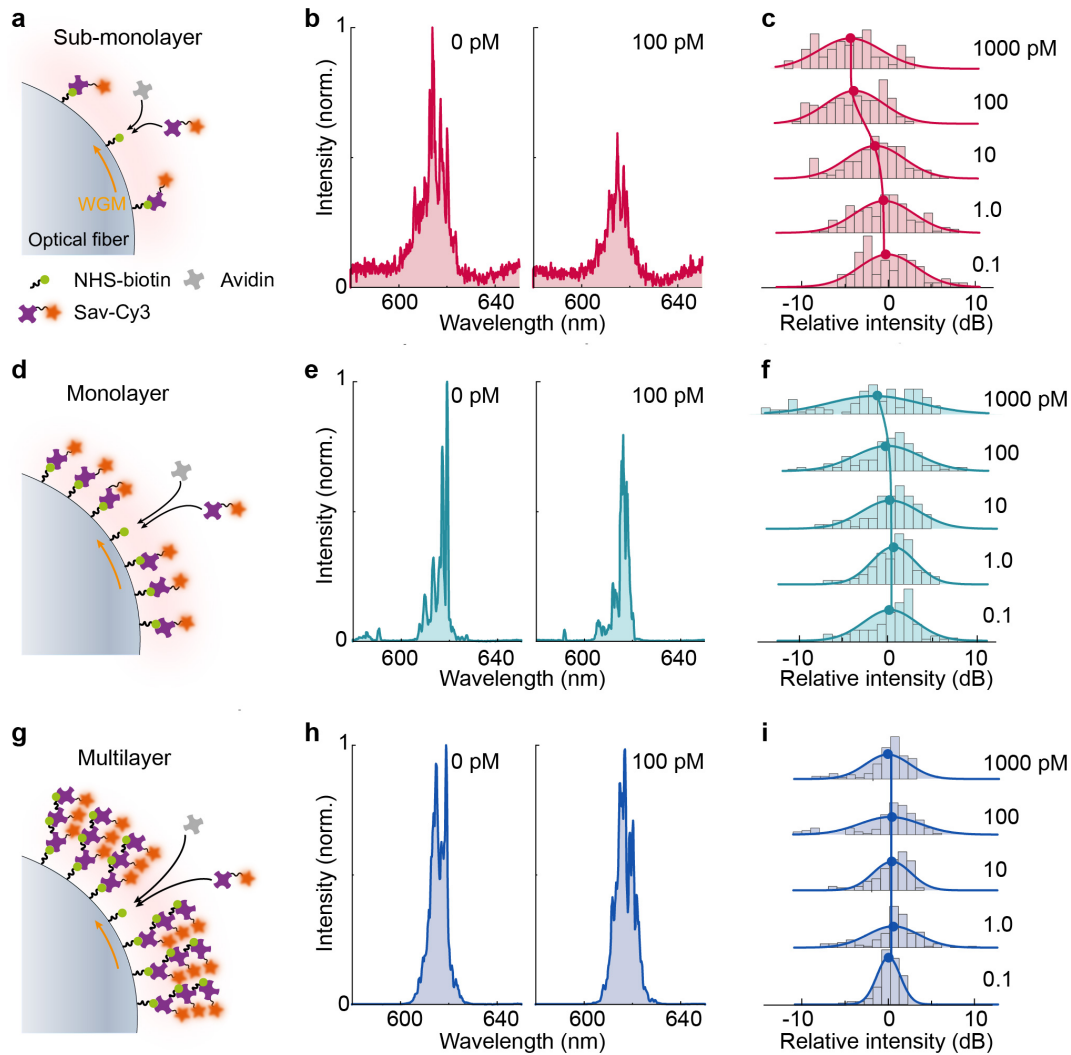

**Figure 3 Comparison of sensitivity of three types of ultrathin biolasers.** Conceptual illustrations of sub-monolayer biolaser (**a**), monolayer biolaser (**d**), and multilayer biolaser (**g**). Comparison of spectra (**b,e,h**) of three types of biolasers with 0 pM and 100 pM of avidin. (**c,f,i**) The dependence of statistical distribution of the laser intensity on the avidin concentration.

We explored the ultimate LOD by reducing the gain molecules down to a threshold density (~ $3.2 \times 10^{-13}$ mol/cm$^2$, Point C in Supplementary Fig. 5). The emission of the sub-monolayer biolasers was recorded with avidin concentration at aM and fM level. No significant change in the laser emission at 0, 10 aM, and 100 aM of avidin, while the laser was extinguished at 1 fM and above (Supplementary Fig. 8). We calculated the relative intensity of all the measurements (yellow curve in Fig. 4), which indicates the capability of distinguishing a step between 100 aM and 1 fM. Compared with the monolayer biolasers, the sub-monolayer biolaser shows a LOD improved by six orders of magnitude (Fig. 4).

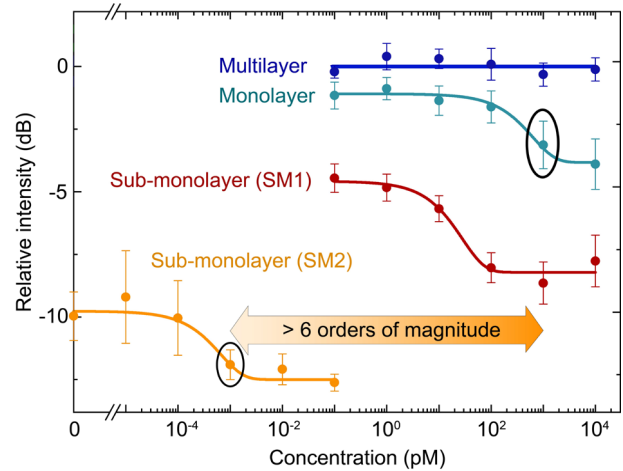

**Figure 4 The ultimate sensitivity of the sub-monolayer biolasers.** SM1 denotes the sub-monolayer biolasers in Fig. 3a and SM2 denotes the sub-monolayer biolasers with gain molecules down to the threshold density.

### 2.4 Detection of Parkinson's disease biomarker

Parkinson's disease is one of the most common neurodegenerative disorders of the elderly and has affected more than 10 million people worldwide[56]. The pathology of PD has a strong correlation with a presynaptic neuronal protein, alpha-synuclein (α-syn)[57,58]. Recent research has indicated that the α-syn in cerebrospinal fluid and blood can be used as a PD biomarker[59]. To explore its capability for medical diagnostics, the sub-monolayer biolasers were designed to detect α-syn with ultrahigh sensitivity (Fig. 5a). The α-syn molecules were sandwiched between the capture antibodies ($Ab_1$) and detection antibodies ($Ab_2$). The immuno-relevant Sav-Cy3 were conjugated to biotin-labeled detection antibodies and thus a stronger laser intensity corresponds to more biomarker molecules. The process of integrating the sandwiched immunoreaction on the sub-monolayer biolaser can be found in Methods.

Firstly, we detected the α-syn in the phosphate buffered saline (PBS) by the sub-monolayer biolasers. The initial surface density of Cy3 molecules was kept at the threshold density, similar to that in Supplementary Fig. 8, so it can lase just above the threshold. Note that α-syn detection is operated in a positive slope mode, and avidin detection is in an inverse sensitivity mode. As predicted by the numerical simulations (Fig. 2c), the laser can achieve an ultimate sensitivity with a sparse conjugating of gain molecules close to the threshold density. The precoating of a portion of Cy3 molecules, as a DC bias, before immunoassay is important to improve the LOD as it directly reduces the required antigen-affiliated gain molecules to reach the laser threshold. The response of the laser emission to the biomarker concentration was calibrated by utilizing the statistical analysis (Supplementary Figure 9), indicating a continuous increase in laser intensity with a higher biomarker concentration. The broadening of the statistical distribution was probably caused by the non-specific binding at higher α-syn concentrations.

Then we calibrated the sensing performance of sub-monolayer biolasers by the detection of α-syn in human serum. Because of its complex components, immunoassay in serum is more challenging, which requires high specificity and high sensitivity. The fiber-supported sub-monolayer biolasers can be fabricated in batches to enable disposable tests (Fig. 5b). Similar to the results in buffer, the statistical distribution in serum shifts towards higher laser intensity with a higher α-syn (Fig. 5c). Figure 5d illustrates the calibration curves obtained in PBS and serum, both of which show good linearity over $R^2 = 98\%$ on the semi-log scale. The lower LOD was estimated to be 0.43 pM and 0.32 pM in buffer and serum, respectively. The LOD in serum is slight lower than that in buffer, as the nonspecific binding of serum components on optical fiber blocks part of the binding sites of the capture antibody, thus reducing the intensity fluctuations. The nonspecific binding phenomenon can also be reflected by the lower intensity and narrower statistical distribution in serum than that in buffer. The disposability and ultrahigh sensitivity of the sub-monolayer biolaser may potentially be applied for healthcare applications in clinics, or hospitals.

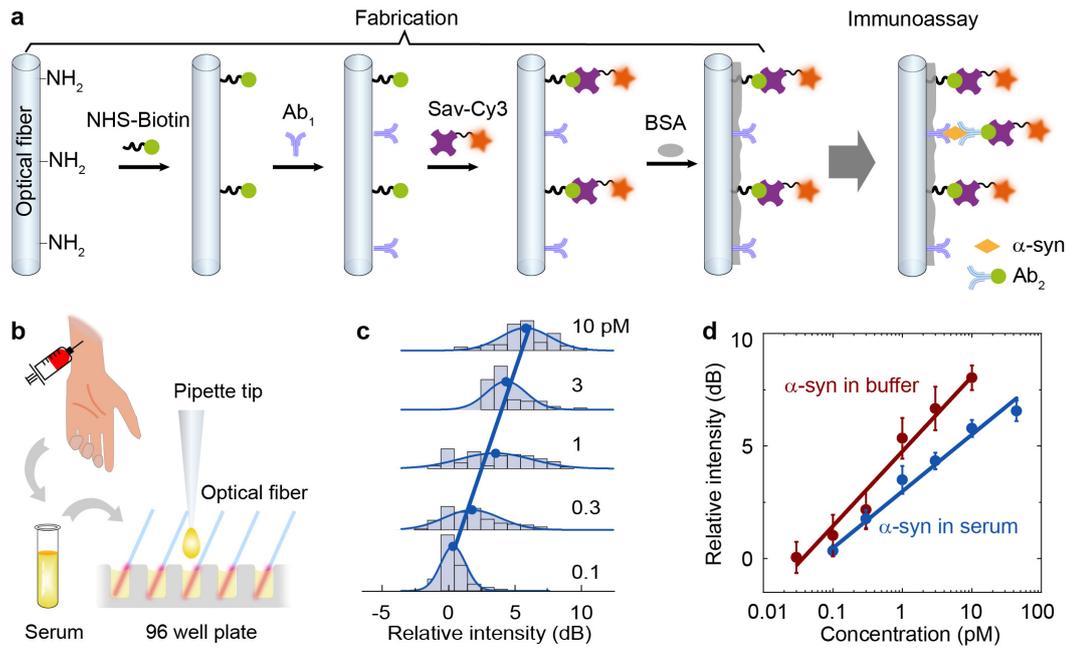

**Figure 5 Alpha-synuclein detection by the sub-monolayer biolaser.** (**a**) Conceptual illustration of bioconjugation protocol for the sub-monolayer biolaser based immunosensor. (**b**) Schematic diagram of massive and disposable immunoassay in serum. (**c**) Statistical distribution of the relative intensity at different α-syn concentrations in serum. (**d**) The calibration curves for α-syn detection in buffer (red) and serum (blue), respectively. Each error bar is calculated from around 60 lasers and denotes the 95% confidence interval.

## 3. SUMMARY

We have developed the sub-monolayer biolasers to perform bioassays with both ultrahigh sensitivity and disposability. The telecom optical fibers have been utilized to provide continuously distributed optical microcavities with an ultrahigh Q-factor of $10^6$, significantly reducing the required gain molecules for lasing. The concept of sub-monolayer biolaser has been confirmed by observing the transition from the stable monolayer biolasers to the gradient sub-monolayer biolasers. The ultrahigh sensitivity has been achieved for protein detection, owing to the greatly enhanced light-matter interaction by the optical resonance in the microcavity and the laser amplification. In addition, the massive production of this laser-based biosensor has been demonstrated, e.g., over one million biosensors (~ 3 cm) can be fabricated in batches from a commercial spool of (50 km) telecom optical fibers. Good reproducibility of the biolasers and their biosensing has been verified experimentally, which enables their disposable uses. Ultrasensitive detection of a PD biomarker in buffer and serum has been demonstrated, offering a general technique for the early diagnosis of major diseases.

## 4. METHODS

### Experimental setup

The telecom single-mode optical fiber (Corning, SMF28e) was exploited by using its silica cladding boundary as the continuously distributed microring resonators (Fig. 1a). As the refractive index of optical fiber is higher than water, the optical fiber supports WGMs on the silica-liquid interface, providing the optical feedback for lasing. The silica microresonators have low roughness, owing to the melt drawing process, and facilitate a high Q-factor for optical resonance. The details of the experimental setup are illustrated in Supplementary Fig. 10. A pulsed laser (Continuum, 532 nm, 5 ns pulse width) was focused by a cylindrical lens into a pump strip of 150 μm × 5 mm. The pump energy density was kept at 1.5 mJ/mm$^2$ with the pump strip perpendicular to the fiber axis. To eliminate the photobleaching effect of the dye molecules, we scanned the pump along the fiber axis with a step of 250 μm and employed a single-pulse pump at each location. The emission spectrum was recorded by a diffraction grating spectrometer (Andor, SR500i) with a thermoelectrically-cooled CCD (Andor, iDus 420A). A blazed grating with 300 lines/mm was used for spectra monitoring. A long-pass filter was used to eliminate the residual pump laser in the detection arm.

### Silanization of optical fibers

The polymer coating of the optical fibers was removed after immersing the fiber in acetone for 1 h. The bare optical fibers were hydroxylated in batches with freshly prepared piranha solution (a 3:7 volume mixture of 30% $H_2O_2$ and 98% concentrated $H_2SO_4$) overnight. The hydroxylated optical fibers were cleaned three times by immersing in deionized (DI) water, each for 5 min. After being washed in dry acetone for 20 min and dried in air, the optical fibers were silanized with APTES (5% in acetone, v/v) for 6 h. The silanized optical fibers were cleaned in acetone, ethanol, and phosphate buffer (PBS, pH = 7.4), each for 10 min. Then, the silanized optical fibers were immersed in PBS and were ready for further experiments.

### Fabrication of sub-monolayer biolasers

The NHS-biotin stock solution with a concentration of 29 mM was prepared by dissolving the lyophilized powder (Aladdin, N103916) in dimethyl sulfoxide (DMSO). The working solution was freshly prepared by diluting the stock solution in PBS. The working solution with a concentration of 200 μM was used for experiments in Figs. 3b to 3c, while a lower concentration of 32 μM was used for exploring the ultimate sensitivity of the sub-monolayer biolasers in Fig. 4 (yellow curve). The silanized optical fibers were incubated in NHS-biotin working solution for 30 min. After washing three times in PBS, each for 10 min, the biotinylated optical fibers were immersed in avidin solutions with various concentrations (0 pM, 0.1 pM, 1 pM, 10 pM, 100 pM, and 1000 pM) for 20 min. After washing, the optical fibers were incubated in 100 μg/ml of Sav-Cy3 (Sigma, No. S6402) in PBS for 40 min to enable the conjugation between Sav-Cy3 and biotin molecules on the optical fiber. After being washed three times with wash buffer (0.05% Tween20 in PBS, v/v), each for 10 min, the optical fibers were immersed in PBS for the laser experiment.

Fabrication of monolayer biolasers

The process of fabricating a monolayer biolaser is the same as the sub-monolayer biolaser, except that a higher concentration of the NHS-biotin working solution (1000 μM) was used for biotinylation.

Fabrication of multilayer biolasers

The multilayer biolasers were fabricated by conjugating multiple layers of the Cy3 molecules on the monolayer biolasers. The monolayer biolasers were immersed in DMSO for 5 min to remove water molecules on the fiber surface and were treated with NHS-biotin working solution (1000 μM) for 30 min. This step was to introduce biotin molecules on streptavidin molecules. Then, the NHS-biotin treated monolayer biolasers were immersed in 100 μg/ml of Sav-Cy3 in PBS for 40 min. The NHS-biotin and Sav-Cy3 treatments were repeated five times to enable five-layer conjugation of Cy3 molecules. After being washed three times with wash buffer, each for 10 min, the optical fibers were immersed in PBS for the laser experiment.

Threshold characterization in a single-use mode

The typical laser threshold curve is shown in Supplementary Fig. 3. Each point was obtained by pumping one location on optical fiber with a single pulse and then renewing the optical fiber location by a scanning step of 250 μm.

Q-factor measurement

The bare optical fibers were treated with piranha solution overnight and followed by washing with DI water. Then, the optical fibers were immersed in PBS for Q-factor measurements ([Supplementary Fig. 2a](#)). An optical spectrum analyzer with the ultrahigh spectral resolution was constructed by a narrow-linewidth tunable laser (New Focus, Model TLB6704-P), a photodetector (New Focus, Model 1801), and a digital oscilloscope (YOKOGAWA, Model DLM3034). The tunable laser was coupled into and out of the optical fiber microcavity through a fiber taper, which was aligned perpendicularly to the optical fiber and finely adjusted through five-dimensional translation stages.

Microscopic characterization of the ultrathin biomolecular film

The fluorescence image of the sub-monolayer biolaser was characterized by a laser confocal microscope (A1R MP$^+$, Nikon), which is equipped with a 20× water immersion objective lens and an excitation wavelength of 561 nm.

Quantify the laser intensity

We calculated the spectral integral of intensity by using $I = \int_a^b i(\lambda)\,d\lambda$. Here, $i(\lambda)$ denotes the spectral distribution of the laser emission. [a, b] defines the spectral range of laser emission. Then, the spectral integral of intensity was converted into decibels by using $I_{dB} = 10\log_{10}(I)$.

Disposability test

Six segments of optical fibers randomly selected from a 10km spool were used for disposable test. For each segment of fiber, the pump laser scanned along the fiber axis with a step of 250 μm, and the laser spectrum at each location was recorded. The statistical result of the sub-monolayer laser emission was shown in [Fig. 2e](#). Then, we calculated the average intensity of the $i$th optical fiber that is denoted as $I_i$. The intensity variation of different optical fibers was calculated by using $\delta = \sigma/\bar{I}$. Here, $\sigma$ is the standard deviation and $\bar{I}$ is the mean value of $I_i$ ($i = 1, 2, \ldots 6$).

Immunoassay with sub-monolayer biolasers

The conceptual illustration for α-syn immunoassay is given in [Fig. 5a](#). The silanized optical fibers were treated with NHS-biotin working solution (32 μM) for 30 min and immersed in DMSO for 5 min to remove water molecules on the fiber surface. The optical fibers were further treated with 50 mg/ml DSS in DMSO for 2 h. After being washed in DMSO for 10 min to remove the residual DSS molecules, the optical fibers were incubated in 120 μg/ml of the capture antibody in PBS for 2 h. It was followed by three times of 5 min wash by the wash buffer, which was exploited for the subsequent wash processes after each incubation. Then, the optical fibers were treated with Sav-Cy3 (100 μg/ml in PBS) for 40 min in order to add a group of background gain molecules. The optical fibers were immersed in the blocking buffer (0.25% BSA in PBS, R&D Systems, DY995) for 1 h. After 5 min washing three times with wash buffer, the optical fibers were immersed in PBS and are ready for immunoassay.

For PD biomarker assay, the buffer solution (1% BSA in PBS) was freshly prepared by dilute the reagent diluent concentrate (×10) with DI water. The analyte solution was freshly prepared by diluting the α-syn stock solution to the desired concentrations in the buffer solution. For PD biomarker detection in serum, the serum (Xinfan Biotechnology, No. 20211119) was diluted by 10 times with PBS. Then, the analyte solution was freshly prepared by diluting the α-syn stock solution with serum. The optical fibers were incubated in α-syn with different concentrations for 1 h. The optical fibers were then immersed in 1.5 μg/ml of the detection antibody for 1 h. We employed 100 μg/ml of Sav-Cy3 in PBS to treat the optical fiber for 40 min. The sub-monolayer biolasers were immersed in PBS for further testing. The capture antibody, α-syn, and detection antibody used in our experiment are from a commercial ELISA kit (R&D Systems, No. DY1338).

Calculating the limit of detection

The relative intensity in Fig. 5 shows a linear relationship with the α-syn concentration on the semi-log scale. Hence, the linear fitting of the calibration curve can be written as $Y = kX+b$, with $k$ and $b$ denoting the slope and intercept of the linear fitting, respectively. $X = \log_{10}(x)$ is the common logarithm of the α-syn concentration ($x$). The LOD is calculated by using $LOD = 10^{(3E_r-b)/k}$. Here, $E_r = 2 \times (1.96 \times SD/\sqrt{N})$ defines the width of the 95% confidence interval in blank control. $SD$ denotes the standard deviation and $N$ denotes the number of data points.

**Acknowledgment**

This work is supported by the National Natural Science Foundation of China (Grant Nos. 61875034, 11825402, 62105006); the 111 Project (B14039); the Sichuan Science and Technology Program (2021YJ0101), the Fundamental Research Funds for the Central Universities (ZYGX2021YGCX007). S.-J.T. is supported by the China Postdoctoral Science

Foundation (Grant Nos. 2021T140023 and 2020M680187).

**Author contributions**

C. G., Y. X. and Y. G. conceived the idea. C. G., X. Y., S. T., Q. Z. and Y. G. designed and carried out the experiments. C. G. and S. T. completed the theoretical analysis and numerical simulations. C. G., X. Y., S. T., Y. W., Y. L., Y. X. and Y. G. contributed to the data interpretation. Y. X., Y. R. and Y. G. supervised the project. C. G., X. Y., S. T., Y. X. and Y. G. completed the manuscript drafting. All authors provided critical feedback and shaped the manuscript.

## 1. Surface modification

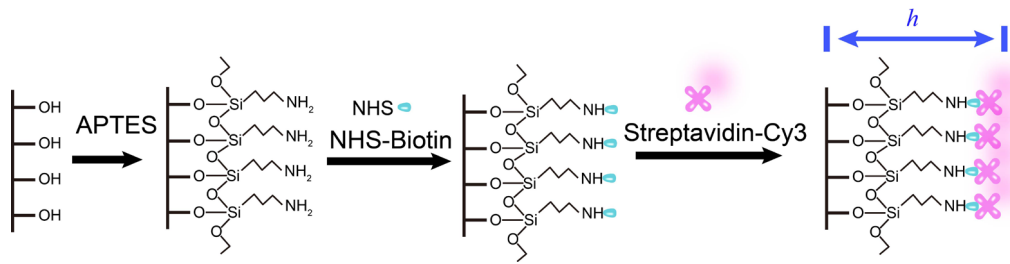

**Supplementary Figure 1.** Illustration of the surface modification for the sub-monolayer biolaser based on the biotin-streptavidin conjugation. $h$ denotes the distance from the Cy3 molecules to the surface of the silica optical fiber.

## 2. Q-factor measurement of the optical fiber microcavities

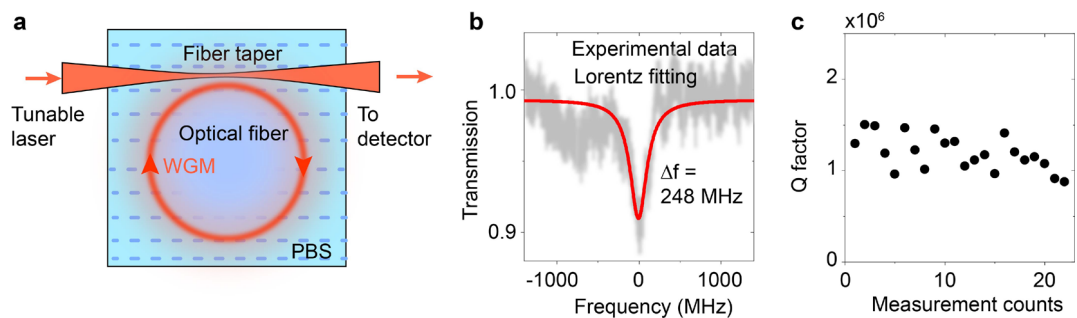

**Supplementary Figure 2. a**, Schematic illustration of the experimental setup for Q-factor tests. **b,** A typical resonant dip in the frequency domain. **c,** The measured Q-factor of the microcavities at different locations of an optical fiber.

## 3. Lasing threshold

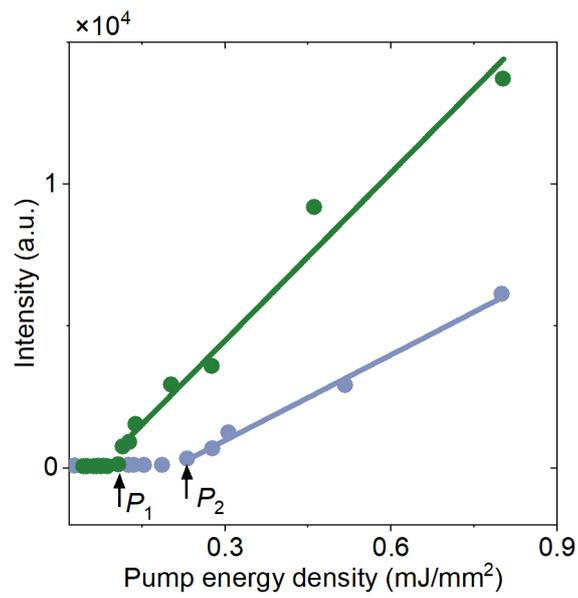

**Supplementary Figure 3.** The threshold curves with biotin concentrations of 400 μM (green) and 200 μM (blue), respectively. Laser thresholds: $P_1 = 0.09$ mJ/mm$^2$; $P_2 = 0.22$ mJ/mm$^2$.

## 4. Numerical simulation of optical resonance in the fiber microcavity

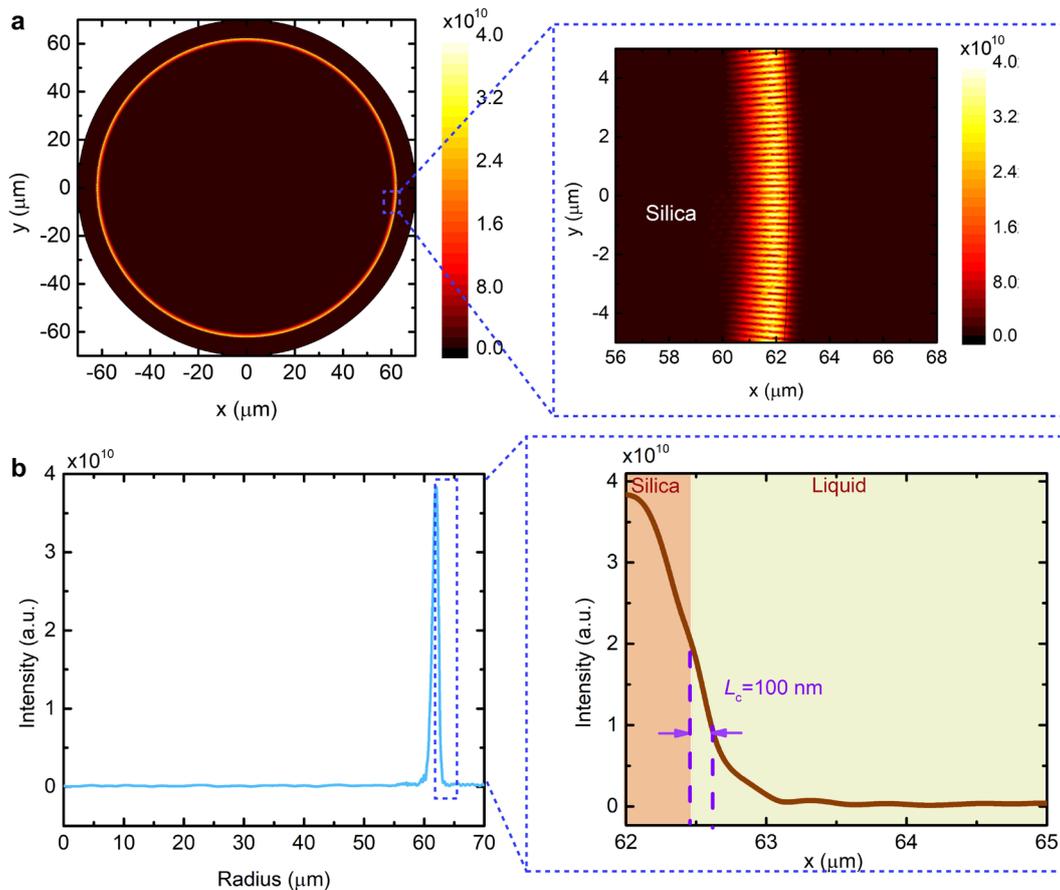

**Supplementary Figure 4. a**, Numerical simulation of intensity distribution on the cross-section of the optical fiber. Inset, enlargement of the region in the dashed box. **b,** The radial intensity distribution with a resonant peak, data was derived from **a**. Inset, enlargement of the

## 5. Analysis of the surface density of Cy3 molecules

The threshold condition for a four-level laser system can be expressed as

$$\eta D_1 \sigma_e(\lambda) = \eta \sigma_a(\lambda) D_0 + \frac{2\pi n}{\lambda_L Q}. \quad (1)$$

Here, $\eta$ = 2.67% is the fraction of energy that interacts with the gain molecules, which can be calculated in Supplementary Fig. 4. $D_0$ and $D_1$ are the density of dye molecules in the ground state and the lowest excited singlet state, respectively. $\sigma_e$ and $\sigma_a$ represent the emission and absorption cross-sections, respectively. $\lambda_L$ denotes the laser wavelength and $Q$ stands for the Q-factor. $n$ is the effective refractive index of the exciting mode of the fiber ring resonator. Thus, the fraction of molecules in the excited state can be given by

$$\gamma = \frac{D_1}{D_0 + D_1} = \frac{\sigma_a(\lambda)}{\sigma_e(\lambda)} \left( 1 + \frac{2\pi n}{\lambda_L D \eta Q \sigma_a(\lambda)} \right). \quad (2)$$

Here, $D = D_0 + D_1$ denotes the density of total gain molecules. According to the rate equation for a four-level laser system, $\gamma$ can also be approximated as

$$\gamma = \frac{w_p \tau_{rad}}{1 + w_p \tau_{rad}} \quad (3)$$

Here, $w_p = I_{th}\sigma_a/E_0\Delta t$ is the normalized pump intensity. $I_{th}$ is the laser threshold. $E_0 = hc/\lambda_p$ is the photon energy of the pump. $\Delta t$ is the pulse width of the pump laser. $\tau_{rad}$ is the lifetime of the excited state.

According to the conclusions in Supplementary Figs. 1 and 4, all the Cy3 molecules conjugated on the optical fiber are located within the reach of the evanescent field and can thus be involved in the optical resonance. Therefore, once the laser threshold is measured, $N$ can be numerically calculated by using Eqs. (2) and (3). The surface density of gain molecules is calculated using $S=D*d$, with $d$ =100 nm denoting the penetration depth (Supplementary Fig. 4). We calculated the dependence of $I_{th}$ on $S$ with various Q-factors and the results are illustrated in Supplementary Fig. 5, through which we can estimate the surface density of gain molecules on fiber by the given threshold.

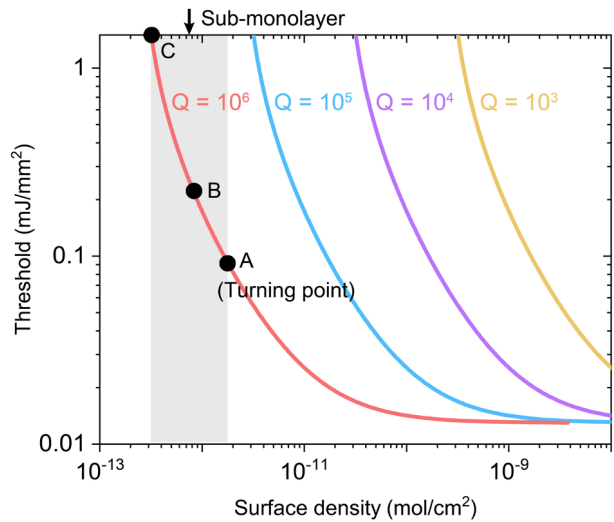

**Supplementary Figure 5** Numerical results of the laser threshold as a function of the surface density of Cy3 molecules. The grey area denotes the sub-monolayer biolaser.

## 6. Reproducibility of lasing threshold

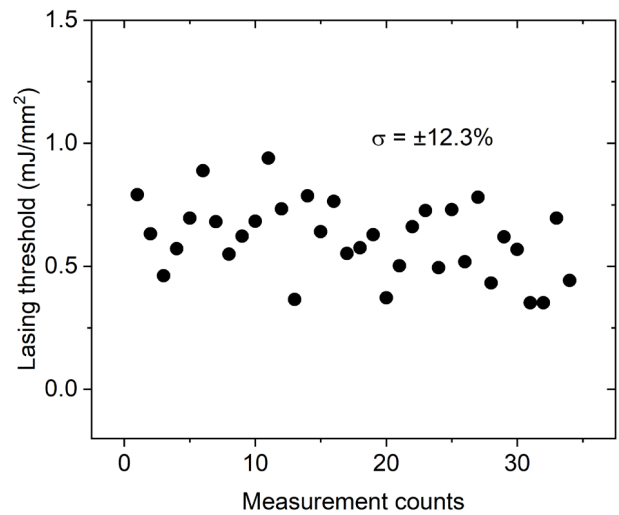

**Supplementary Figure 6.** Lasing thresholds from different sub-monolayer biolasers. These sub-monolayer biolasers was fabricated using 100 μM NHS-biotin.

## 7. Statistical distribution of laser emission

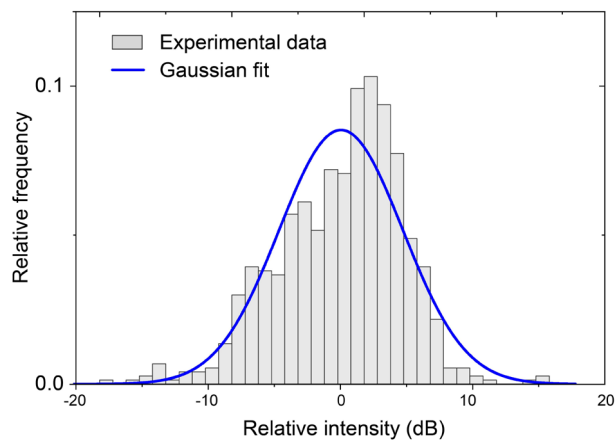

**Supplementary Figure 7.** Frequency histogram of relative intensity. Data are extracted from 736 sub-monolayer biolasers.

## 8. Theoretical model for sensitivity analysis

The laser intensity can be written as

$$I_{laser} = A\left(\frac{I_{pump}}{I_{th}} - 1\right). \tag{4}$$

Here, $I_{pump}$ is the pump intensity and $I_{th}$ is the laser threshold. $A$ is a constant.

According to the rate equation for a four-level laser system, the fraction of molecules in the excited state $\gamma$ can be approximately expressed as

$$\gamma = \frac{w_p \tau_{rad}}{1 + w_p \tau_{rad}}, \tag{5}$$

with

$$w_p = \frac{I_{th}}{E_0 \Delta t} \sigma_a. \tag{6}$$

Here, $w_p$ denotes the normalized pump intensity. $E_0 = hc/\lambda_p$ is the photon energy. $\Delta t$ is the pulse width of the pump laser. $\tau_{rad}$ and $\sigma_a$ are the lifetime on the excited state and absorption cross-section of Cy3 molecules, respectively.

Combining with Supplementary Eqs. (1) to (3), we derive that

$$I_{laser} = A\left[\frac{I_{pump}\sigma_a \tau_{rad}}{E_0 \Delta t}\left(\frac{1}{\gamma} - 1\right) - 1\right]. \tag{7}$$

This equation shows the dependence of the laser intensity on $\gamma$.
By combining Supplementary Eqs. (6) and (4), we can calculate the laser emission as a function of the surface density ($S$).

In immunoassay, the number of gain molecules is proportional to the number of conjugated antigen molecules. According to Supplementary Eq. (4) and Eq. (6), the sensitivity of the laser-based sensor to the antigen on fiber can be defined as

$$\frac{dI_{laser}}{dS} = \frac{dI_{laser}}{d\gamma} \cdot \frac{d\gamma}{dD} \cdot \frac{dD}{dS}. \tag{8}$$

Here,

$$\frac{dI_{laser}}{d\gamma} = -A \cdot \frac{I_{pump}\tau_{rad}\sigma_a}{E_0 \Delta t} \cdot \frac{1}{\gamma^2}, \tag{9}$$

$$\frac{d\gamma}{dD} = -\frac{2\pi n}{\sigma_e \lambda_L \eta Q} \cdot \left(\frac{S}{d}\right)^2, \tag{10}$$

and

$$\frac{dD}{dS} = \frac{1}{d} \tag{11}$$

According to the analysis above, we theoretical calculation of the sensitivity as a function of Cy3 surface density was shown in Fig. 2c. The results indicate an increase in sensitivity with a lower surface density and an ultrahigh sensitivity can be achieved when the surface density decreases to the threshold density.

Comparatively, in the fluorescence tests, the emission intensity can be written as

$$I_{FL} = B \cdot S \cdot I_{pump} \qquad (11)$$

Here, $B$ is a constant. The sensitivity of the fluorescence-based sensor can be defined as

$$\frac{dI_{FL}}{dS} = B \cdot I_{pump} \qquad (12)$$

Therefore, the sensitivity of the fluorescence-based sensor remains independent of the molecular surface density.

## 9. Exploring the ultimate sensitivity of the sub-monolayer biolasers.

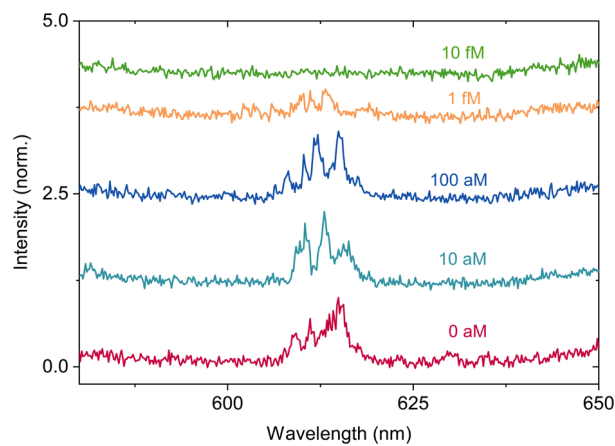

**Supplementary Figure 8.** Spectral evolution of sub-monolayer biolaser with different concentrations of avidin.

## 10. Alpha-synuclein detection in buffer

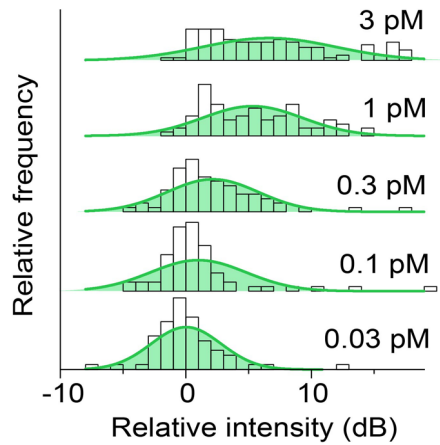

**Supplementary Figure 9.** Statistical distribution of the relative intensity at different α-syn concentrations in buffer.

## 11. Experimental setup

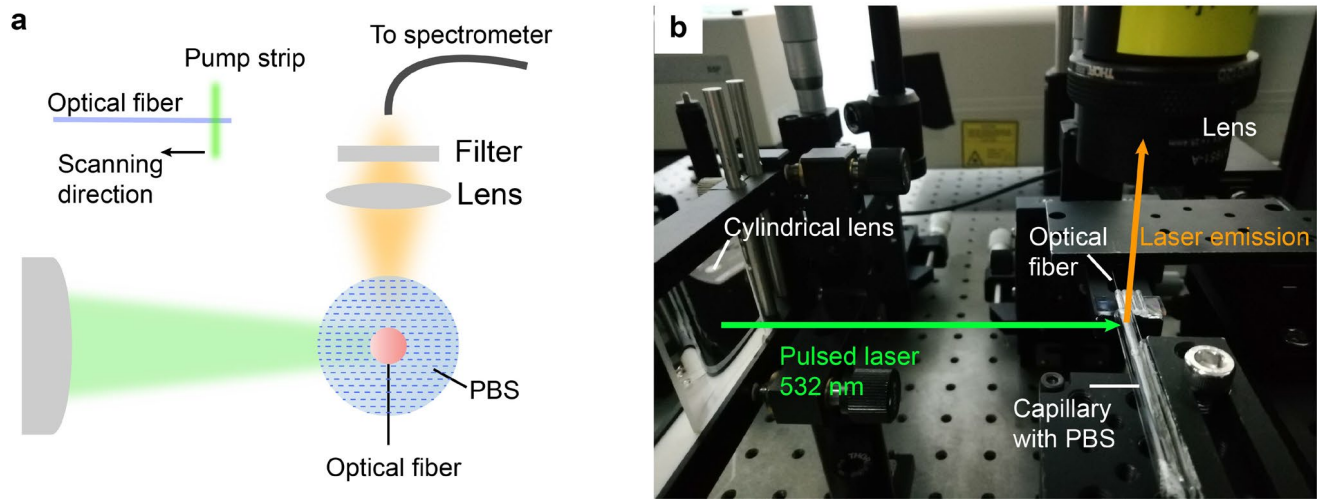

**Supplementary Figure 10.** Conceptual illustration (a) and the photo (b) of the experimental setup.